\documentclass[a4paper,12pt,reqno,superscriptaddress]{revtex4}
\usepackage[centertags]{amsmath}
\usepackage{amsfonts}
\usepackage{amssymb}
\usepackage{amsthm}
\usepackage{graphicx}
\usepackage{newlfont}
\usepackage{stmaryrd}
\usepackage{mathrsfs}
\usepackage{euscript}
\usepackage{rotating}

\usepackage{color}

\theoremstyle{plain}

\theoremstyle{definition}

\theoremstyle{remark}

\numberwithin{equation}{section}

\newcommand{\opunit}{\text{1}\kern-0.22em\text{l}}

\DeclareMathAlphabet{\mathpzc}{OT1}{pzc}{m}{it}

\begin{document}

\title{\large{Saturation of front propagation in a reaction-diffusion process\\ describing plasma damage in porous low-$k$ materials.

}}

\author{Soghra Safaverdi}
\affiliation{Instituut voor Theoretische Fysica, K.U.Leuven,
Belgium}
\author{Gerard T. Barkema}
\affiliation{Theoretical Physics, Utrecht, The Netherlands }
\affiliation{Instituut-Lorentz for Theoretical Physics, Leiden University, The Netherlands}
\author{Eddy~Kunnen}
\affiliation{imec, Kapeldreef 75, 3001 Heverlee, Belgium}
\author{Adam M. Urbanowicz}
\affiliation{Semiconductor Physics Section, Department of Physics and Astronomy, K.U.Leuven, Belgium}
\author{Christian Maes}
\affiliation{Instituut voor Theoretische Fysica, K.U.Leuven, Belgium}

\keywords{front propagation, Stefan problem, reaction-diffusion system}

% --------------------------------------------------------------

\begin{abstract}
We propose a three--component reaction--diffusion system yielding
an asymptotic logarithmic time-dependence for a moving interface.
This is naturally related to a Stefan-problem for which both one-sided
Dirichlet-type and von Neumann-type boundary conditions are considered.
We integrate the dependence of the interface motion on diffusion and
reaction parameters and we observe a change from transport behavior
and interface motion $\sim t^{1/2}$ to logarithmic behavior $\sim \ln t$ as a function of time.
We apply our theoretical findings to the propagation of carbon depletion
in porous dielectrics exposed to a low temperature plasma.
This diffusion saturation is reached after about 1 minute in typical
experimental situations of plasma damage in microelectronic fabrication.  We predict the general
dependencies on porosity and reaction rates.
\end{abstract}

\maketitle

\section{Introduction}
The generation and propagation of interfaces in reaction-diffusion systems
is of great importance in a variety of scientific and technological
contexts.  In nature, the interface often spatially separates two
macroscopically different homogeneous regimes in which one phase or
product dominates another one. Typically, at least one particle-type
diffuses through the medium while reacting with other particles. Births
and deaths depend on the spatial location and on the presence of other
particle-types.  Then, when installing appropriate initial and boundary
conditions one often sees the appearance of an interface and one asks
for the typical temporal behavior in its initial growth and further
propagation. The case we treat here is boundary driven from one side,
with symmetric bulk diffusion for one particle type that invades the
region of a second particle type, while constantly being annihilated
by yet a third type of particles.  That creates a moving phase boundary
and thus goes under the general name of Stefan problem \cite{stef,gup}.

The present work is inspired by experimental investigations on plasma
damage in porous low-capacitive materials (low-$k$, see \cite{vice}).  One imagines here a porous
SiOC material which is exposed via one of its physisorbedaries to radicals
such as active oxygen atoms.  These radicals diffuse through the material
while removing hydrophobic carbon-hydrogen groups and while being
chemi-adsorbed in the pores of the material.  One then observes the
appearance and motion of an interface separating a rich (pristine)
carbon-phase from a damaged (oxygen-rich) layer. It is important here
to understand the time-dependence of that motion and its dependence
on diffusion and reaction parameters, which can further be related to
material and chemical properties.  We come back to this application in
the separate section \ref{app}.

 Of course, the literature on interface motion from reaction-diffusion
equations is vast, and many different phenomena have been reported.  In particular, damage spreading and related front
propagation can take many different forms which we cannot at all review here; see e.g.~\cite{gra,bord,genrd} for general research aspects.
  As just one very famous example
for one-dimensional front propagation, as we also have
here, reaction-diffusion equations such as of the type of
Kolmogorov-Petrovsky-Piskounov show interface motion between two
homogeneous phases at constant speed \cite{kpp}.  That then takes the
form of travelling waves in which one phase advances over the other phase.
Here, however, the proposed model shows the motion of the interface $x(t)
\simeq \ln (1+at + b\sqrt{t})$ (in one dimension as function of time $t$)
to be logarithmic which means that after an initial period diffusion
saturation occurs.
The main point is that the diffusive motion is
boundary driven and that the propagation is limited by the reactions.
For the purpose of the application, the logarithmic behavior is caused
by the disappearance of the oxygen radicals before reaching the moving
interface separating them for the carbon-rich environment.  The $\sqrt{t}$
within the logarithm, mostly visible in an initial time-period, originates
directly from the diffusive motion.

The next section contains the specific mathematical model in the form
of coupled partial differential equations.  They give a mean-field
description of the reaction-diffusion system.  The main finding is the
change from a diffusive to a specific logarithmic time-dependence for
the interface $x(t)$ as above, with the study of dependencies on initial
conditions and dynamical parameters.  We give a physical heuristics for
understanding the basic characteristics via a linearized model.\\

The section \ref{num} on numerical results summarizes the behavior of our
model, as a function of the physical parameters. Our results agree with
the analytical results of the linearized model, but the higher flexibility of simulations
allow for obtaining additional information and predictions concerning
the dependence on the various parameters in the model.

Section \ref{app} goes deeper into the specific application that triggered
our study, that of the propagation of damage in porous low-$k$ materials.
We recall the main aspects and the meaning of our result for that context.
In particular, reference \cite{ime} contains already the basic set of differential equations, but no systematic analysis
was performed and the distinction between different (one-sided) boundary conditions was not considered.
The present paper also adds the interesting connections with reaction-diffusion systems of Stefan type.

\section{Model and main findings}

There are three types of particles.  The first are called free
O-particles, which can diffuse but can also change into bound O-particles
and further annihilate with either bound O-particles or with C-particles:
\[
\mbox{free } O \overset{\alpha}{\rightarrow} \mbox{ bound } O,
\;\;
\mbox{ free } O + \mbox{ bound } O \overset{\kappa }{\rightarrow} \emptyset,
\;\;
\mbox{ free } O + C \overset{R}{\rightarrow} \emptyset
\]
and no other reactions or diffusions take place.  The letters are chosen in the context of the later application
in Section \ref{app} with O for oxygen and C for carbon.  The word ``free'' will there be understood as ``reactive'' and ``bound''
will mean adsorbed to the inner surface, either physically or chemically.  Once oxygen atoms are absorbed on the surface they start to move around until they recombine and desorb.  In other words, since in the experimental conditions desorption of bounded oxygen molecules typically occurs through recombination, reversible desorption is not taken into account, \cite{des,kun}.
For the moment we continue to use the more general words ``free'' and ``bound,'' as the mathematical set-up can have different natural realizations.\\
The dynamics is
further determined by the initial condition at time zero when there
are only C-particles with homogeneous density, and by the boundary
conditions.  The physics we have in mind concerns a three-dimensional
region of locations $(x\geq 0,y,z)$, which is isotropic over the
$(y,z)-$coordinates.  For the boundary conditions at the $x=0$
plane we consider two possibilities.  The first, of Dirichlet-type,
specifies a constant density of free O-particles at $x=0$; the second,
of von Neumann-type, gives a constant flux of particles entering the
region at $x=0$.  For the questions and the problem at hand we restrict
ourselves to an effective mean field treatment, ignoring fluctuations
or microscopic inhomogeneities, and the geometry is one-dimensional.

We consider the coordinate $x\geq 0$ in which to express the profile
of particle densities as a function of time $t\geq0$.  There are
thus three types of concentrations $\rho(x,t), c(x,t)$ and $m(x,t)$
respectively for free O-particles, C-particles and bound O-particles.
The reaction-diffusion system is then defined via the equations
\begin{eqnarray}\label{dif}
\dot{\rho} &=& D\,\rho'' - \kappa \,\rho\,m - R\,c\,\rho - \alpha\,\rho\nonumber\\
\dot{c} &=& - R\,c\,\rho \nonumber\\
\dot{m} &=& - \kappa \,\rho\,m + \alpha\,\rho
\end{eqnarray}
where the dotted left-hand sides are time-derivatives.  The concentration
$\rho$ (of the free O-particles) undergoes a second spatial derivative
in the right-hand side of the first line of \eqref{dif}; the parameter
$D>0$ being the diffusion constant.  The other parameters $\kappa, R$
and $\alpha$ are positive reaction rates.  The initial condition $(t=0)$ specifies
\begin{equation}
c(x,t=0) = 1, \;\; \rho(x,t=0) = m(x,t=0) \equiv 0, \mbox{ for } x > 0\label{inic}
\end{equation}
meaning that we start with C-particles fully occupying the material.
On the other hand, the boundary conditions ($x=0$) can be taken either
\begin{equation}\label{bc}
\mbox{ Dirichlet-type: } \rho(x=0,t) = V,\;\; \mbox{ \ or von Neumann-type: } \rho'(x=0,t) = - J
\end{equation}
where we can take $V=1$ without loss of generality; the $J>0$ stands for the rate of free O-particles entering the region which for the moment we leave
as a parameter.
It seems more natural to opt for von Neumann-type boundary conditions
for the specific application we have in mind, but we checked that
after a transient time there is really no difference between Dirichlet
and von Neumann conditions.  In the case of von Neumann-type boundary
conditions, the value $\rho(x=0,t)$ converges from being zero at time
zero to a fixed finite value (equal to some $V>0$ depending on $J$) exponentially fast as time evolves.
%Figs.~1 versus 2 illustrate the equivalence between Dirichlet- and von
%Neumann-type boundary conditions for the propagation of the interface.
As a consequence, we mostly concentrate on just one type, that of
Dirichlet boundary conditions with $\rho(x=0,t)=1$ for all times, but
further discussion on possible differences will follow below.

The results of numerical integration of the differential equations
\eqref{dif} are detailed in Section \ref{num}.  Our main finding is
that the concentration $c(x,t)$ of C-particles (defining the absence
of damage in the application of Section \ref{app}) shows an interface
$x(t)$ separating a C-poor from a C-rich phase.  There are {\it a priori}
a number of different possible mathematical definitions of the position
of the interface $x(t)$. A first one is to solve $c(x(t),t) \simeq 1/2$
(or some other number) for $x(t)$, which certainly
makes sense when the interface is sharp enough, after some transient time,
but is less useful for very small times.  A second definition uses the same conditions as above but we add the equation
$\dot{x}(t)=-D\,\rho'(x(t),t)$ for the interface speed.  That brings the problem in the natural neighborhood of Stefan problems
%is to work with the concentration $\rho$ of free O-particles and to find $x(t)$ so
%that $\rho(x(t),t) \simeq 0$ with
%interface speed given by $\dot{x}(t)=-D\,\rho'(x(t),t)$ which brings the problem in the natural neighborhood of Stefan problems,
\cite{stef,gup}, also valid for very small
times but more involved numerically.  Whatever (reasonable) definition
we take for $x(t)$ we invariably find that the motion is logarithmic of
the general form $x(t) = A\ln(1 + at + b\sqrt{t})$ in a typical range of
parameters for diffusion and reaction rates $D, \kappa, R$ and $\alpha$.
The coefficients $A,a$ and $b$ depend on these parameters, see also in
Section \ref{num}.  For small times the motion is diffusive $x(t) \sim
\sqrt{t}$ while it saturates as $x(t) \sim \ln t$ for large times $t$.
The transition between the two regimes is basically determined by the
amount of reactivity, the more the O-particles can be annihilated,
the faster saturation sets in.

An approximate treatment towards the second Stefan-like definition of the interface motion described above can already be made visible for
a much simpler linear system to which we turn next. We simplify
\eqref{dif} to a one-component reaction-diffusion equation by imagining
an autonomous linear dynamics for $\rho$ of the form
\begin{equation}\label{heet}
\dot{\rho}(x,t) = D\,\rho'' - \beta\,\rho
\end{equation}
The equation \eqref{heet} mimics the first line of \eqref{dif} in
the approximation $\kappa \,m + R\,c + \alpha = \beta > 0$ constant,
which we expect to be mostly reasonable when $c=0$ and
$m = $constant which makes sense in the damaged region ($0<x<x(t)$) where the carbon was depleted.  The right boundary denoted by $\ell$ in what follows therefore plays the role of interface.  We also can think of $\beta$ as a measure
of reactivity and interaction, while $D$ measures the diffusion. The main advantage is of course that now the
equation \eqref{heet} can be solved exactly with boundary conditions
$\rho(x=0,t)=1, \rho(x=\ell,t)=0$ and initially $\rho(x,t=0)=0$.
The solution is
\begin{eqnarray}\label{linsol}
\rho(x,t) &=&\rho_\ell(x) + e^{-\beta t}\sum_1^\infty a_n\,e^{-\frac{D\,n^2\pi^2}{\ell^2}t}\,\sin\frac{n\pi}{\ell}x,\,\;\mbox{ where} \nonumber\\
\rho_\ell(x) &=& \frac 1{1-e^{2k\ell}}e^{kx} + \frac 1{1-e^{-2k\ell}}e^{-kx},\;\; k\equiv\sqrt{\beta/D}\nonumber\\
&=&  -\sum_1^\infty a_n\sin \frac{n\pi}{\ell}x \;\;\mbox{ which determines the } a_n
\end{eqnarray}
and $\rho_\ell$ is the stationary solution on $[0,\ell]$.  In fact we
see that $\rho(x,t)$ converges exponentially fast to $\rho_\ell$ with
rate $\beta>0$ uniformly in $\ell$.  We can already learn a great deal
from the behavior at the edge $x=\ell$.  It is easy to find that
\begin{equation}\label{sinh}
\rho_\ell'(\ell) = \frac{-k}{\sinh k\ell}
\end{equation}
We think now of $\ell$ as the position $x$ of the interface and
$-\rho_\ell'(\ell)$ is then the speed of the interface, so that Stefan's
condition $\dot{x}(t) = D\,k/\sinh(k x(t))$ can be integrated to give
\begin{equation}\label{xsol}
x(t) = \sqrt{\frac{D}{\beta}}\,\ln[1 + \beta t +\sqrt{2\beta t + \beta^2t^2}]
\end{equation}
We clearly recognize the behavior $x(t) \propto \sqrt{D/\beta}\,\ln t$
for large times $t$ resulting in a straight line in Fig.~1 and intersecting the vertical axis (small $t$) at $\sqrt{D/\beta}\,\ln (2\beta)$.    For small times $t$ \eqref{xsol} gives $x(t) \propto \sqrt{t}$, and $x(t=1)\simeq \sqrt{2D}$ for small $\beta>0$.  We can thus find the effective $D$ and $\beta$ for
\eqref{heet} from the intersections of the $x(t)$ as function of $\ln t$ with the vertical axis, as in Fig.~1, which provide consistency checks
for our modeling equations.\\

Similarly, we can also solve \eqref{heet} with von Neumann-type boundary
condition~$\rho'(x=0,t)=-J$ but still putting $\rho(x=\ell,t)=0$ and
initially $\rho(x,t=0)=0$.  The solution is
\begin{eqnarray}\label{linsol}
\rho(x,t) &=&\rho_s(x) + e^{-\beta t}\sum_1^\infty b_n\,e^{-\frac{D\,(n-1/2)^2\pi^2}{\ell^2}t}\,\cos\frac{(n-1/2)\pi}{\ell}x,\,\;\mbox{ where} \nonumber\\
\rho_s(x) &=& \frac{J}{k}[\frac{e^{-kx}}{1 + e^{-2k\ell}} - \frac{e^{kx}}{1+e^{2k\ell}}],\;\; k\equiv\sqrt{\beta/D}\nonumber\\
&=&  -\sum_1^\infty b_n\cos \frac{(n-1/2)\pi}{\ell}x \;\;\mbox{ which determines the } b_n.
\end{eqnarray}
Exponentially fast in time $t$, $\rho(0,t) = J/k + O(e^{-k\ell})$, and
hence we find that for $J\sim\sqrt{\frac{\beta}{D}}$ the von Neumann
condition imitates the Dirichlet condition with fixed $\rho(0,t)=1$
and $\rho_\ell(x)\simeq \rho_s(x) \simeq \exp -kx$ for large $\ell$, at least
for $\beta\neq 0$.
Clearly then, when the rate $\beta$ would go to zero, some differences between Dirichlet and von Neumann
(one-sided) boundary conditions can be expected, as we indeed recover in Section \ref{infres} for the real model.  The next section
goes back to the full nonlinear model equations \eqref{dif} but we will find the many general features to be similar to our chosen approximation above.

\section{Numerical results}\label{num}

We now turn to the full set of equations \eqref{dif}, and switch to
computer simulations. First, we study the system for a specific set of
parameters $D=1.0$, $\kappa=0.1$, $R=1.0$ and $\alpha=0.1$, and apply
Dirichlet boundary conditions; the primary motivation for this specific
choice of the parameters lies in their numerical convenience. We apply
a forward Euler integration scheme to equations \eqref{dif} with a time
step of $\Delta t=0.001$, and store the density profiles of C-particles,
free and bound O-particles at times $t=1,2,\dots$.   Based on our analytic
solution \eqref{xsol} of the simplified description \eqref{heet}, we expect
that for long times the interface position $x(t)$, defined as the
(linearly interpolated) location where the C-particle density equals 1/2,
changes in time as $x(t)\approx A\ln(1 + at)$, for some values of $A$
and $a$ which are determined by the system parameters $D,\kappa,R$ and
$\alpha$. Fig.~1 shows that this expected long time behavior is confirmed by the
simulation data.

\begin{figure}[h]
\centerline{\includegraphics[height=10cm,angle=270]{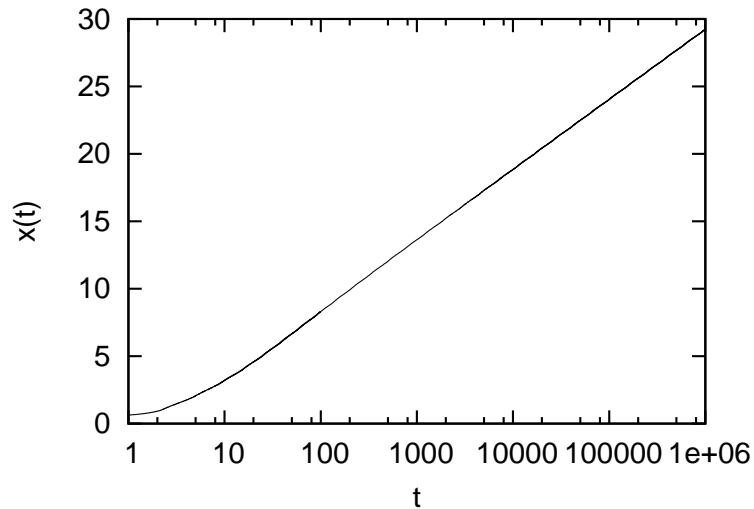}}
\caption {Position of the interface $x(t)$ as a function of time.
For longer times, the curve approaches a straight line, indicating that
asymptotically $x(t)=A\ln(1+at)$.}
\end{figure}

In the same simulations, we also determined the concentrations of C-particles
and free O-particles as a function of position, for the
times $t/100=1, 2, 4, 8, \dots, 8192$. These are $c(x,t)$ respectively $\rho(x,t)$
in equations \eqref{dif}.  The resulting profiles of C-concentration
are shown in Fig.~2a. The C-density profile attains a sigmoid-like
shape, which gradually moves deeper into the material. As the times of
the curves are separated by a factor of two, and the depth increases
with the logarithm of time, the curves are more or less equidistant.

The concentration profile of free O-particles is shown in Fig.~2b.
The general shape is that of an exponentially decaying curve up to the
interface position $x(t)$, after which it decays again exponentially, but
with a much steeper slope. Once the interface has significantly moved
into the material, the gradient of free O-concentration at the surface becomes
constant even though we applied Dirichlet boundary conditions. This confirms
our earlier analytic observation that the behavior for Dirichlet and
von Neumann conditions soon becomes indistinguishable.

\begin{figure}[h]
\centerline{\includegraphics[height=10cm,angle=270]{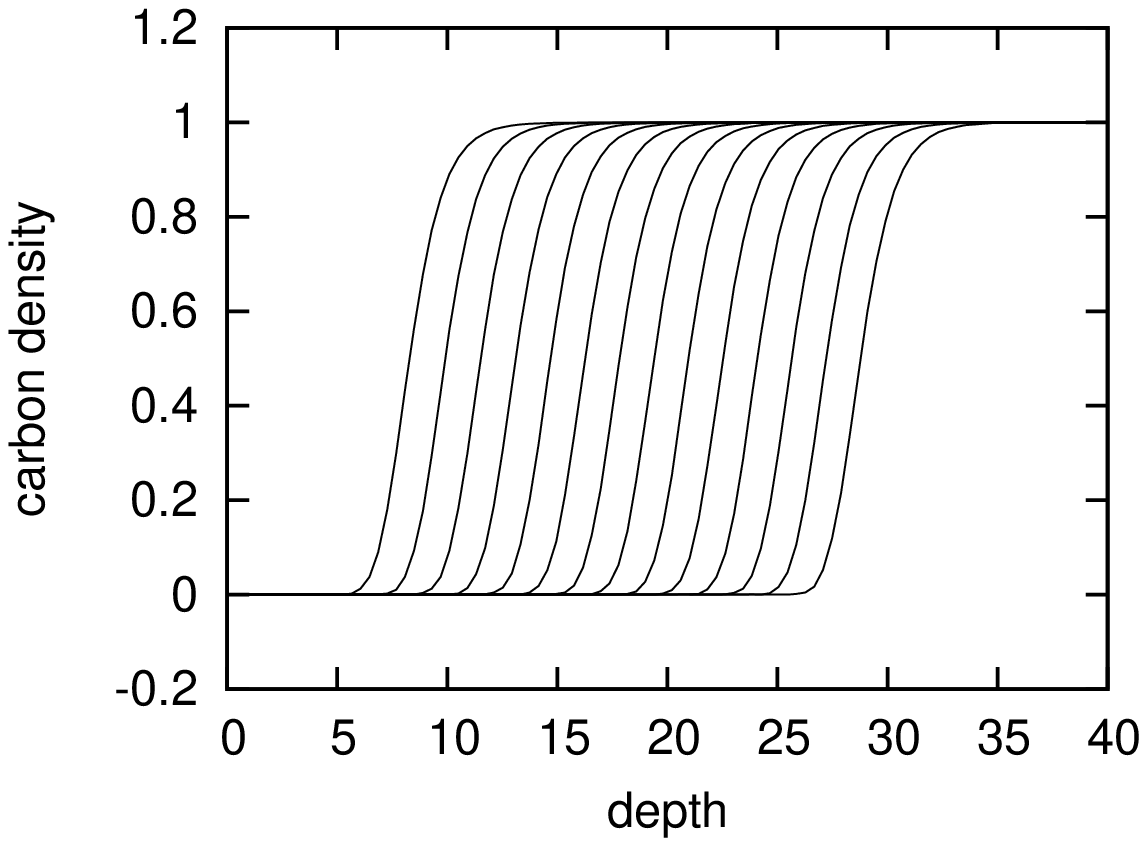}}

\centerline{\includegraphics[height=10cm,angle=270]{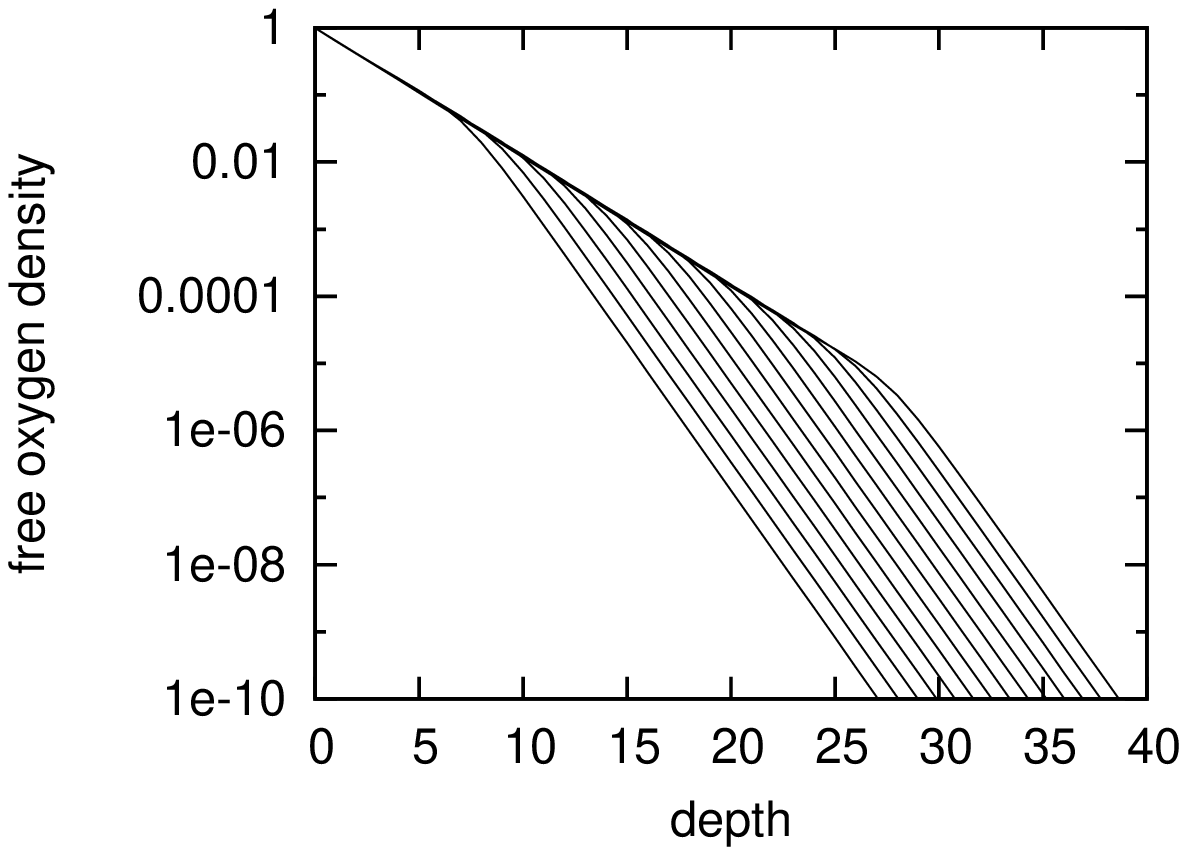}}
\caption {Profile of the C-concentration (top panel) and free O-concentration (lower panel) as a function of depth, at times $t/100=1,2,4,\dots, 8192$.
Note that the top panel has a linear vertical scale, while the lower panel has a
logarithmic vertical scale.}
\end{figure}

\subsection{Influence of diffusion}\label{difro}

An important ingredient is to understand the influence of the diffusion
constant $D$ on the temporal behavior of the interface.  After all,
the porosity and the architecture of the material can modify exactly
that diffusive behavior.

We repeated the simulations with different values for the diffusion coefficient: $D=0.5, 1, 2$ and 4. Fig.~3 shows the rescaled interface
position $x(t)/\sqrt{D}$, indicating that the main effect of variations
in $D$ is a rescaled amplitude: $A\sim \sqrt{D}$,  as expected from
the linear model in \eqref{xsol}.  It appears therefore that we understand
the influence of the connectivity of the porous material on the
time-dependence of the interface.

\begin{figure}[h]
\centerline{\includegraphics[height=10cm,angle=270]{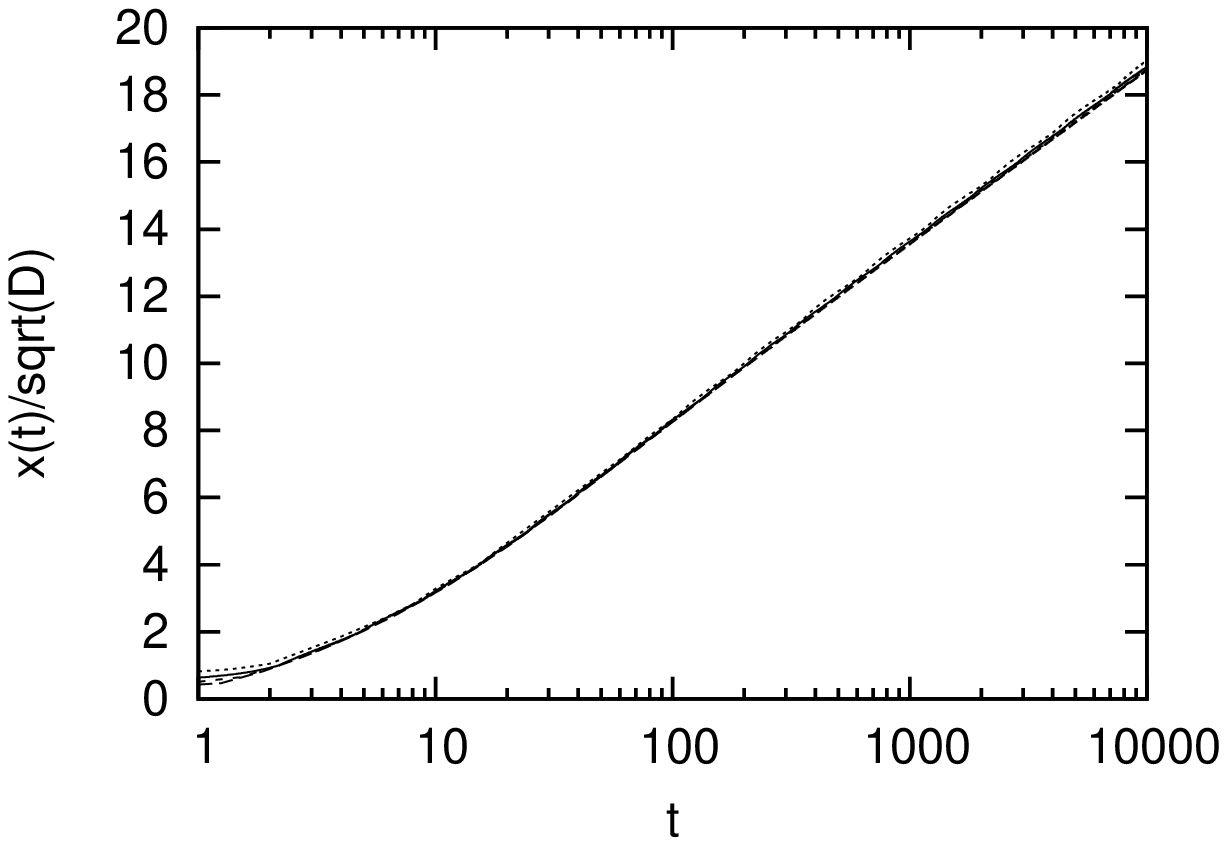}}
\caption {Rescaled position of the interface $x(t)/\sqrt{D}$ as a function of time,
for simulations in which $D=0.5, 1, 2$ and 4.
For longer times, the curves fall on top of each other, indicating that
asymptotically $A\propto \sqrt{D}$.}
\end{figure}

\subsection{Influence of reactivities}\label{infres}

The adsorption of the free O-particles and the chemical annihilation
of the free O-particles with the bound O-particles (e.g. to create a $O_2$ molecule which is chemically neutral for C-particles) and with the
C-particles is the reactive content of our model.  Obviously the
C-particles must remain to define an interface at all, but we can
imagine lowering very much the adsorption of the free O-particles.
In the extreme case we exclude the adsorption of free O-particles by
setting $\alpha=0 $.  Then, we have $m(x,t)=0$ always and everywhere which
simplifies \eqref{dif} somewhat. In the case of Dirichlet type-boundary
condition $\rho(x=0,t) = 1$, we basically have transport of O-particles
through a layer of size $x(t)$ with difference in concentration fixed (and
equal to 1):  we thus expect from Fick's law that the current $\dot{x}(t)
\sim 1/x(t)$ is proportional to the concentration gradient which gives
us $x(t) \sim \sqrt{t}$. We have checked that this behavior is indeed
found for very small adsorption rates, $\alpha=0$ and $\kappa=0$.
Fig.~4 shows the numerical evidence.\\

  On the other hand, with von
Neumann boundary condition $\rho'(x=0,t) = -J$ and for very
small absorption rates $\alpha=0$ and $\kappa=0$, there is
a continuous influx of free O-particles leading to a
concentration of them at the surface which increases in a power law
fashion. That can be understood from the following calculation. Let us take $J=1$ and consider the function
\[
\rho_1(x,t)\equiv \sqrt{t}\,e^{-x/\sqrt{t}}
\]
satisfying indeed $\partial_t \rho_1(x=0,t) = -1$.
Then, $\partial_t \rho_1(x,t) = \rho_1(x,t)/t\;[1 + x/\sqrt{4t}]$ and
$\partial_x^2 \rho_1(x,t) = \rho_1(x,t)/t$, so that $\rho_1$ appears as a solution of the diffusion equation
for $x\ll \sqrt{t}$.  We thus conclude that $\rho_1(0,t)$ grows as $\sqrt{t}$.
 Numerically, we find
that the lack of saturation of free O-concentration
at the exposed surface does not alter the asymptotic behavior of the interface, but it does lead
to a logarithmic correction. The behavior is consistent with $x(t) \sim
\sqrt{t\ln t}$.  Also for von Neumann boundary conditions, numerical
evidence is presented in Fig.~4.

\begin{figure}[h]
\centerline{\includegraphics[height=10cm,angle=270]{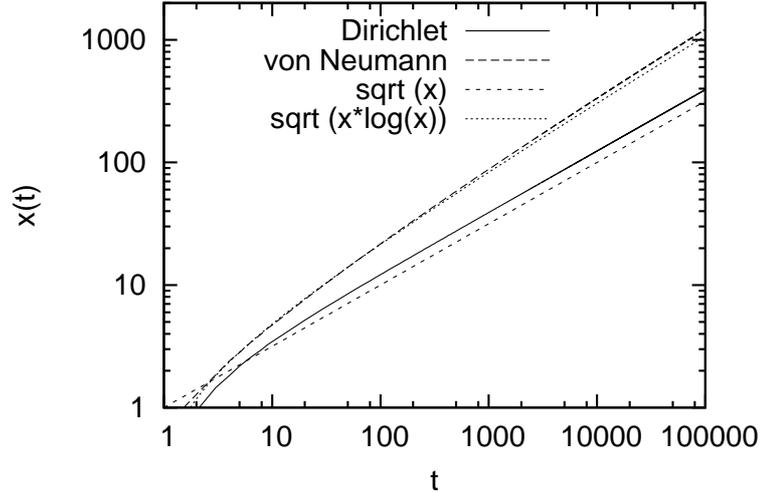}}
\caption {Position of the interface $x(t)$ as a function of time,
for simulations in which the adsorption of free oxygens is excluded.
With Dirichlet boundary conditions, the asymptotic behavior can be well
fitted by $x(t) \sim \sqrt{t}$; with von Neumann boundary conditions,
this becomes $x(t) \sim \sqrt{t \,\ln t}$.}
\end{figure}

\section{Application}\label{app}
Low dielectric constant (low-$k$) materials are typically used in semiconductor manufacturing to compensate for the RC-delay and power consumption increase in microelectronic devices associated with their continuing scaling-down \cite{vice,vol}. The microelectronics community has adopted $k$ as a representation of the relative permittivity or dielectric constant in contrast to the scientific communities using $\varepsilon_r$. Nowadays, porous low-$k$ dielectrics based on silica and silsesquioxanes with 10-15\% of organic hydrophobic groups are the most favoured class of materials for advanced interconnect technology nodes \cite{vol}. The hydrophobic groups are bonded to a Si-atom in the SiO${}_{4/2}$ matrix and can be represented as O${}_3\equiv $SiCH${}_x$. To reach a $k$-value below 3.0, introduction of artificial porosity is needed (the $k$-value of air is close to 1) \cite{vice}. Advanced low-$k$ materials have a porosity of 25-50\% and a typical pore size being close to 2--2.5 nm. Although the matrix of these materials has properties similar to traditional SiO${}_2$, their chemical stability and reactivity strongly depend on porosity. Materials with interconnected pores are chemically active because of the high diffusion rate of active chemicals.  In particular the surface diffusion of adsorbed radicals is much faster than the desorption, which is our motivation not to include reversible adsorption-desorption in our model equations \eqref{dif}, see \cite{mas, mar} for specific experimental estimates.\\

 During device fabrication, the porous low-$k$ dielectrics are exposed to various plasmas that might degrade their properties. The most challenging plasma treatment is related to plasmas based on oxygen- or hydrogen containing chemistries that are used to remove organic photo-resist masks used for pattern transfer from the low-$k$ films \cite{urba}.  The challenge arises from the similar chemical composition of the low-$k$ film
and the organic photo-resist mask. Both materials contain C-H bonds. The plasma radicals that remove the organic mask also penetrate the porous low-$k$ dielectric film destroying its Si-CH${}_3$ bonds. This leads to formation of polar Si-OH groups as a result of direct chemical reaction with active radicals or saturation of  $\equiv$Si$\bullet$ broken bonds by OH groups from ambient moisture.  The Si-OH groups are centers for further moisture adsorption \cite{bak}. The absorbed moisture with a $k$-value of 80 can fill the whole open pore volume of the material significantly increasing its $k$-value which is summarized as damage.\\

  Two commonly known approaches are used for the photo-resist mask removal: (i) a low temperature, low pressure anisotropic plasma, where the photo-resist is removed by an ion-assisted process, oxidizing or reducing plasma chemistries at low temperatures \cite{kun} and (ii) hydrogen-based downstream plasma (DSP) where the resist is removed at high temperatures by a thermally activated chemical process \cite{ada}. According to recent publications \cite{kun,ada}, both approaches remove C from low-$k$ materials. However, ion-assisted processes remove hydrophobic Si-CH${}_3$ groups and graphitized-C (residual-C that remains after film fabrication), while a thermally activated chemical process with hydrogen plasma removes only graphitized-C from low-$k$ dielectrics. In both cases the depth of C-carbon removal can be directly measured with spectroscopic ellipsometry \cite{ime,urba}.  Various experimental results have shown the typical time-dependence of the depth of C-removal. In particular the typical logarithmic time-dependence was observed, as the one we have derived in the present paper.\\

 The depth of the plasma radical penetration in the low-$k$ pores determines the extent of plasma damage. Clearly then, the role of Knudsen diffusion is important and our analysis in Section \ref{difro} clarifies some aspects. For example, the extent of plasma damage can be a few times higher for mesoporous dielectrics with high porosity levels in comparison with nanoporous dielectric with low porosity levels \cite{sha,moo}.\\

As a word of caution and limitations of our approach, plasma damage of low-$k$ materials is a complex phenomenon that results in changes of their bonding structure and pore morphology \cite{vice}.  The complexity also increases due to the fact that porous low-$k$ dielectrics are usually amorphous materials with a random pore structure \cite{vice}.  Our set of reaction-diffusion equations \eqref{dif} can be generalized to include some of these aspects but soon a more detailed simulation analysis becomes necessary.

\section{Conclusions}\label{heu}

The propagation of damage when a pristine material is exposed to diffusing
and reacting particles saturates logarithmically.  The initial position
of the interface is diffusive $\sim \sqrt{t}$ but soon saturates $\sim \ln t$
when reactions prohibit the propagation of the damage.  The problem
can be modeled as a reaction-diffusion system of coupled differential
equations where the position of the interface can be determined by viewing it as a
Stefan problem.  We have discussed Dirichlet and von Neumann boundary conditions, solved a linear approximation and we have found the detailed influence of reactivities
and diffusivity.  Numerical integration of the coupled reaction-diffusion equations is in full accord with experimental findings
 in the context of microelectronic fabrication of low-$k$ materials.\\

\noindent{\bf Acknowledgment:}
We are grateful to Mikhail Baklanov for guidance on the experimental back-ground of the work and for stimulating discussions.

\bibliographystyle{plain}

\end{document}